\documentclass[acus]{JAC2003}

\usepackage{graphicx}
\usepackage{booktabs}
\usepackage{amsmath}
\usepackage{array}

\begin{document}
\title{BEAM DYNAMICS INVESTIGATION OF THE 101.28 MHz IH STRUCTURE AS INJECTOR FOR THE HIE-ISOLDE SC LINAC\\[-3mm]}
\author{M.A. Fraser$^{\dag \ddag}$, M. Pasini$^{\S \ddag}$, R.M. Jones$^{\dag}$, D. Voulot$^{\ddag}$\\
        $^{\dag}$The University of Manchester, Oxford Road, Manchester, UK\\
        $^{\dag}$The Cockcroft Institute, Daresbury, Warrington, Cheshire UK\\
        $^{\S}$Instituut voor Kern- en Stralingsfysica, K.U.~Leuven, Celestijnenlaan 200D, Leuven, BE\\
        $^{\ddag}$CERN, Geneva, Switzerland}

\maketitle

\begin{abstract}
	The first phase of the HIE-ISOLDE project at CERN consists of a superconducting (SC) linac upgrade in order to increase the energy of post-accelerated radioactive ion beams from 2.8~MeV/u to over 10~MeV/u (for $A/q = 4.5$). In preparation for the upgrade, we present beam dynamics studies of the booster section of the normal conducting (NC) REX-ISOLDE linac, focused on the longitudinal development of the beam in the 101.28~MHz Interdigital H-mode Structure (IHS), employing a Combined Zero Degree Structure (KONUS)~\cite{konus}, pulsing at a high gradient of over 3~MV/m. The evolution of the transverse emittance in the SC linac depends critically on the injected longitudinal phase space distribution of particles from the existing linac and, with a better understanding of the beam dynamics upstream, the performance of the upgrade can be optimised. Data taken during the commissioning phase of the REX-ISOLDE linac is analysed to understand the properties of the beam in the booster and combined with beam dynamics simulations which include the realistic fields of the IHS, determined from both simulation and perturbation measurement.
\end{abstract}

\section{THE HIE-ISOLDE INJECTOR}

\begin{figure}[htb]
   \centering
   \includegraphics*[width=80mm]{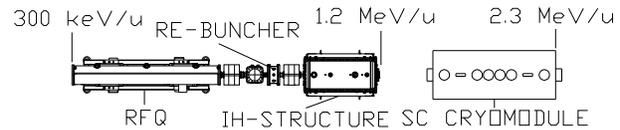}
   \vspace*{-\baselineskip}
   \caption{The layout of the injector and the first low energy SC cryomodule.}
   \label{inj}
\end{figure}

The REX infrastructure which will act as injector to the HIE-ISOLDE linac is shown schematically in Figure~\ref{inj}. The IHS is a compact cavity containing separated regions of acceleration and focusing, including a triplet of quadrupoles and a section of drift tubes operating as a re-buncher, boosting the beam from 0.3 to 1.2~MeV/u over 1.5~m, by providing an effective voltage of up to 4.2~MV~\cite{rex}. The KONUS beam dynamics design constrains the acceptance in both the transverse and longitudinal phase spaces, which can produce severe deterioration in beam quality if the machine is not set at the correct working point. The objective of the investigation was to benchmark the \texttt{LORASR} code, see for example~\cite{lorasr}, used to design the IHS with numerical multi-particle simulations of the realistic fields and to assess the emittance dilution through the booster before injection into the SC upgrade.

\section{MEASUREMENTS AND SIMULATIONS OF THE IHS}

\begin{figure}[htb]
   \centering
   \includegraphics*[width=85mm]{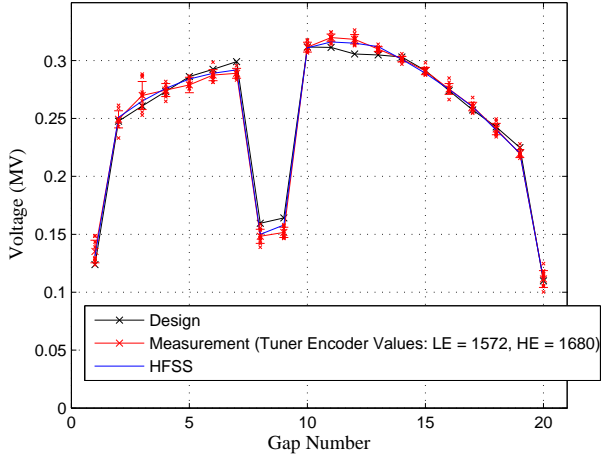}
   \vspace*{-\baselineskip}
   \caption{A comparison of the voltage distribution of the 20 gaps in the IHS, normalised to a total of 5.04 MV.}
   \label{vgaps}
\end{figure}

The accelerating field of the IHS was measured using the bead pull method. An experimental set-up originally designed for measuring the field flatness of Linac 4 structures at CERN was adapted to facilitate a measurement on the REX beam line. The set-up included control software for the bead velocity and data acquisition, see~\cite{setup}. The use of the measured fields in the beam dynamics simulations demands an accuracy similar to that of the mechanical tolerances of the cavity to ensure a correct phasing of the KONUS dynamics. Although the bead velocity was controlled to within an rms stability of 3 \%, a post-processing routine was necessarily developed to allow for the reconstruction of the accelerating field using many measurements. Over repeated measurements the accelerating profile correlated very closely with that extracted from \texttt{HFSS} simulations of the nominal design geometry, compared in terms of gap voltage in Figure~\ref{vgaps}, where a small deviation to the design voltage can be observed. The encoder positions of the tuners were re-calibrated to fixed external references.

\begin{figure}[htb]
   \centering
   \includegraphics*[width=85mm]{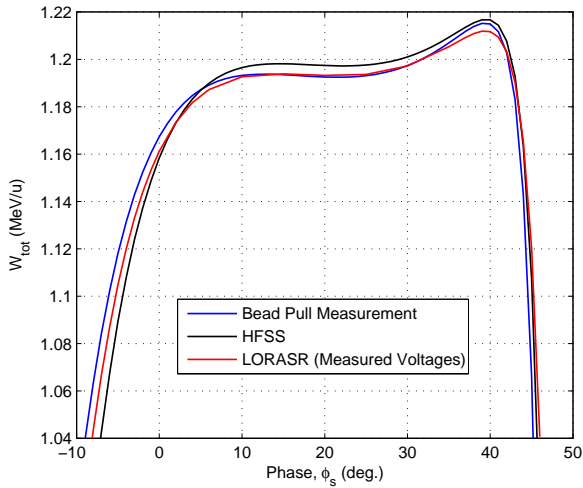}
   \vspace*{-\baselineskip}
   \caption{A code comparison with the phase scan of the ejection energy.}
   \label{w_phi}
\end{figure}

The ejection energy of a single particle is shown in Figure~\ref{w_phi} as its phase is scanned. The calculation was done numerically in both the measured and simulated fields and compared with \texttt{LORASR}, in which the fields are parameterised in simple analytic formulae.
The numerical simulations are in good agreement with \texttt{LORASR} and allow for a calibration of the phase between codes.

\section{THREE GRADIENT EMITTANCE MEASUREMENT}

\begin{figure}[htb]
   \centering
   \includegraphics*[width=85mm]{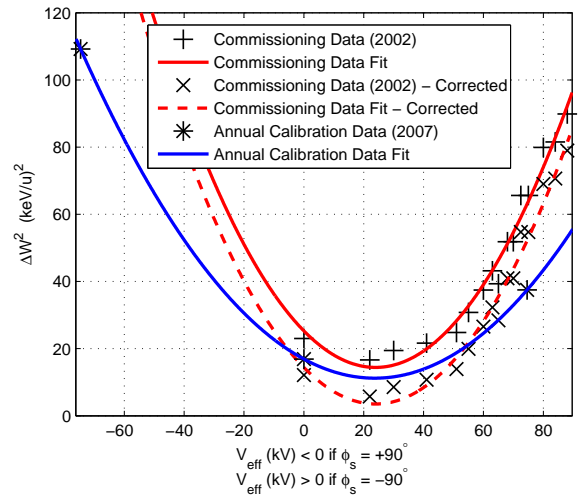}
   \vspace*{-\baselineskip}
   \caption{The data sets fitted with quadratic curves according to the three profile method.}
   \label{fits}
\end{figure}

The longitudinal beam parameters at the exit of the RFQ were reconstructed from data taken during the commissioning of REX in 2002~\cite{se}. A second and much smaller data set from a set-up period in 2007 was also analysed. The data sets represent measurements of the energy spread as a function of the re-buncher voltage operating at the non-accelerating phases of $\pm90^\circ$. Using the three gradient method and representing the buncher as a thin accelerating gap the square of the energy spread $\Delta W^2$ can be expressed as a quadratic function of the effective voltage $V_{eff}$, which is parameterised by the longitudinal Twiss beam parameters in front of the buncher as,

\begin{equation*}
\frac{\Delta W^2}{A^2}  = \epsilon_{0}\left[\left(\frac{q}{A}\right)^2\beta_{0}V_{eff}^2\pm2\left(\frac{q}{A}\right)\alpha_{0}V_{eff}+\gamma_{0}\right],
 \label{fit_twiss}
\end{equation*}
where $A/q$ is the mass-to-charge state and all other symbols have their usual meaning. The three gradient measurement of emittance is analogous to the quadrupole scan method commonly used for measuring the transverse emittance. More details of a similar measurement can be found in~\cite{slac}. There is little information regarding the uncertainty and resolution of the two data sets. We focused our analysis on the more comprehensive data set taken during commissioning. The resolution of the 2007 data set is estimated at 3~keV/u or 1\% by comparing the measured energy spread with the buncher switched off to \texttt{PARMTEQ} simulations of the energy spread after the RFQ,~\cite{sieber}. It is treated as a systematic error and removed from the measured energy spread in quadrature to create the corrected data set. The fitted data sets are presented in Figure~\ref{fits} and the fitted parameters presented in Table~\ref{paras}, along with the 2007 data set corrected for the resolution. The random errors are not plotted but assumed to be equally weighted across the data set at 0.5~keV/u or 0.2 \%.

\begin{figure}[htb]
   \centering
   \includegraphics*[width=85mm]{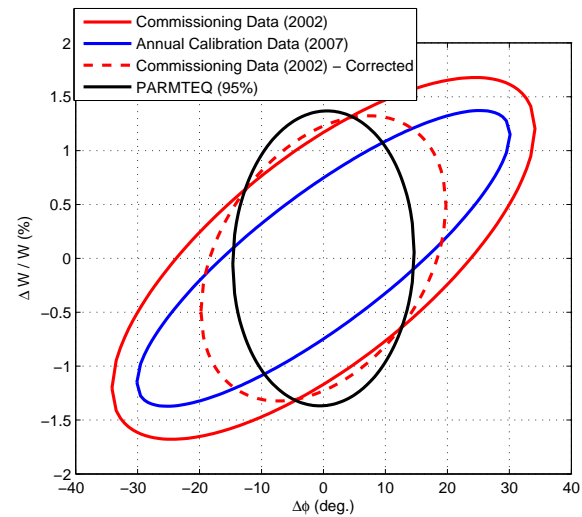}
   \vspace*{-\baselineskip}
   \caption{The longitudinal phase space ellipses reconstructed at the RFQ exit.}
   \label{ellipses}
\end{figure}

\begin{table}[hbt]
   \setlength\tabcolsep{2pt}
   \centering
   \caption{Beam Parameters Reconstructed at the RFQ Exit}\label{paras}
   \smallskip
   \resizebox{\columnwidth}{!}{%
        \begin{tabular}{@{}lcccc@{}}
            \toprule
               \textbf{Parameter}  & \textbf{PARM.} & \textbf{2002} & \textbf{2002 - Corr.} & \textbf{2007}\\
            \midrule
                $A/q$         & 4.5  & 4.5 & 4.5 & 3.5    \\
                $\alpha$         & -0.04  & -0.92 $\pm$ 0.23 & -0.45 $\pm$ 0.15 & -1.41    \\
                $\beta$ (ns / keV/u)        & 0.10       & 0.24 $\pm$ 0.03 & 0.15 $\pm$ 0.04 & 0.33  \\
                $\epsilon_{95\%}$ ($\pi$ ns keV/u)        &    1.8     & 3.3 $\pm$ 0.4  & 2.1 $\pm$ 0.5& 1.9     \\
              \bottomrule
        \end{tabular}
   }
\end{table}

The thin gap approximation for the REX re-buncher, which is in reality a three-gap split-ring cavity, was shown to be valid and less significant than the resolution of the dipole switchyard magnet used to measure the energy spread for effective voltages below 100 kV for $\phi_s = -90$ and below 60 kV for $\phi_s = +90$~\cite{rex_emit}. The effect of emittance growth in the buncher using realistic electromagnetic fields was also shown to be small. The lack of knowledge of the uncertainty and resolution involved in the energy spread measurements limits the conclusions that can be drawn. However, \texttt{LANA} simulations have shown that poor measurement resolution results in larger values of $\alpha$ calculated at the RFQ, see~\cite{rex_emit}. The corrected data set is indeed more consistent with the beam ellipse predicted by \texttt{PARMTEQ}, containing 95 \% of the bunch. A study was undertaken to assess the possibility of improving the measurement. The study concluded that a resolution of 0.18 \% is in theory possible using a 1 mm slit in front of the dipole and a quadrupole after, to ensure the spectrometer operates as a point-to-point system, and that the beam parameters could be reconstructed to within 15 \%. It is therefore viable to repeat the above measurements to check the discrepancy with simulation. Although such a procedure is tedious with the diagnostics presently available it is foreseen to introduce solid-state technology for time and energy domain measurements.

\section{MULTIPARTICLE SIMULATIONS}

\begin{figure}[htb]
   \centering
   \includegraphics*[width=85mm]{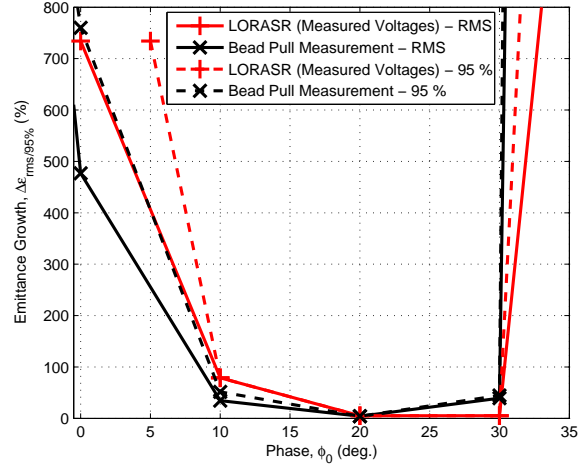}
   \vspace*{-\baselineskip}
   \caption{The longitudinal emittance growth as a function of phase of the bunch centre in the middle of the first gap, comparing \texttt{LORASR} with simulations of the measured fields.}
   \label{emit}
\end{figure}

In order to understand the longitudinal beam stability along the injector 1000 particles were simulated in the measured accelerating field with a total emittance of 2 $\pi$ keV/u ns, consistent with the results in Table~\ref{paras}. The beam was convergent on entry to the IHS as required for beam stability. The narrow phase acceptance of the IHS is highlighted in Figure~\ref{emit} indicating that the longitudinal emittance growth can be kept below 10 \%, consistent with \texttt{LORASR}.

\section{CONCLUSION}

A benchmarking of the \texttt{LORASR} longitudinal dynamics was presented using the numerical integration of the equations of motion in the measured accelerating field of the REX-IHS. The longitudinal emittance growth was shown to be less than 10 \% at an injected emittance of 2 $\pi$ keV/u ns; an emittance level which itself was shown to be consistent with the analysis of commissioning data using the three profile method. The measurement could be repeated with a resolution five times smaller. The three profile measurement provides a method for tuning the injector which would ensure that the beam is correctly matched into the acceptance of the IHS and that its working point is set accurately. A solid-state diagnostic system is foreseen to make this quick and reliable.

\section{ACKNOWLEDGEMENTS}

M.A. Fraser acknowledges the receipt of funding from the ISOLDE Collaboration Committee and the Cockcroft Institute. We would like to express our gratitude for the help and support of Giovanni De Michele and Piero Posocco.


\begin{thebibliography}{9}
\bibitem{konus}
U. Ratzinger, The IH-structure and its capability to accelerate high current beams, PAC '91.
\end{thebibliography}
\end{document}